\documentclass[draftcls,onecolumn,letterpaper,peerreview]{IEEEtran}
\makeatletter
\let\NAT@parse\undefined
\makeatother
\usepackage{mathrsfs,amssymb,amsmath}
\usepackage[ps2pdf, colorlinks=true, citecolor=blue, bookmarks=true, pdfstartview=FitH, bookmarksnumbered=true, bookmarksopen=true, hypertexnames=false, breaklinks=true]{hyperref}
\newcommand{\SC}{\mathscr{C}}
\newcommand{\SB}{\mathscr{B}}
\newcommand{\T}[1]{\tilde{#1}}
\newcommand{\tr}{\textrm{Tr}(\SC)}
\newcommand{\im}{\textrm{Im}_\SB (\SC)}
\newcommand{\qm}{\textrm{GF}(q^m)}
\newcommand{\qq}{\textrm{GF}(q^2)}
\newcommand{\q}{\textrm{GF}(q)}
\newcommand{\p}{\textrm{GF}(p)}

\newtheorem{thm}{Theorem}
\newtheorem{cor}[thm]{Corollary}
\newtheorem{lem}[thm]{Lemma}
\newtheorem{prop}[thm]{Proposition}
\begin{document}
\title{Self-orthogonality of $q$-ary Images of $q^m$-ary Codes and Quantum Code Construction}
\author{Sundeep~B~and~Andrew~Thangaraj,~\IEEEmembership{Member,~IEEE}
\thanks{Sundeep B. and A. Thangaraj are with the Department of Electrical Engineering in the Indian Institute of Technology Madras, Chennai, India.}}

\maketitle
\begin{abstract}
A code over GF$(q^m)$ can be imaged or expanded into a code over GF$(q)$ using a basis for the extension field over the base field. The properties of such an image depend on the original code and the basis chosen for imaging. Problems relating the properties of a code and its image with respect to a basis have been of great interest in the field of coding theory. In this work, a generalized version of the problem of self-orthogonality of the $q$-ary image of a $q^m$-ary code has been considered. Given an inner product (more generally, a biadditive form), necessary and sufficient conditions have been derived for a code over a field extension and an expansion basis so that an image of that code is self-orthogonal. The conditions require that the original code be self-orthogonal with respect to several related biadditive forms whenever certain power sums of the dual basis elements do not vanish. Numerous interesting corollaries have been derived by specializing the general conditions. An interesting result for the canonical or regular inner product in fields of characteristic two is that only self-orthogonal codes result in self-orthogonal images. Another result is that image of a code is self-orthogonal for all bases if and only if trace of the code is self-orthogonal, except for the case of binary images of 4-ary codes. The conditions are particularly simple to state and apply for cyclic codes. To illustrate a possible application, new quantum error-correcting codes have been constructed with larger minimum distance than previously known.
\end{abstract}
\IEEEpeerreviewmaketitle
\section{Introduction}
Linear codes are subspaces of vector spaces over a finite field. To find efficient codes over a particular field, it is often-times beneficial to look for codes over an extension field. Since the extension field is a vector space over the base field, any vector in a vector space over the extension field can be {\it imaged} into a vector over the base field by expanding each coordinate with respect to a basis for the extension field. Reed-Solomon (RS) codes, one of the most successful codes in practice, form a popular example of a code construction over extension fields. The binary image of RS codes is used in many applications such as magnetic hard disk drives, optical drives and deep space communications. Codes formed as images of a code over an extension field turn out to have some useful properties and advantages such as protection against burst errors and ease of encoding and decoding.

While images of codes have been successfully used in practice, a precise description of their algebraic properties has been a challenge in the field of coding theory for a long time. Problems related to codes over extension fields and their images continue to remain unsolved today \cite[Chapter 10]{sloane}. A few problems have attracted some attention in the past. The problem of determining when the $q$-ary image of a cyclic code over $\qm$ is cyclic was solved in \cite{seguin} by using a module structure for images. Perhaps the most interesting problem related to images is the determination of minimum distance of the image of a code. Many versions of this problem have been studied in works such as \cite{sakakibara, retter}. Properties of the images of codes have also been studied with respect to soft-decision decoding \cite{lacan, vardy}. 

In this paper, we study the problem of self-orthogonality of $q$-ary images of $q^m$-ary codes ($q=p^r$, $p$ prime). We derive necessary and sufficient conditions on the original code and the basis such that the image is self-orthogonal with respect to a given product. Our primary result is that self-orthogonality of the image with respect to a particular product (such as $\sum xy$) depends on self-orthogonality of the original code with respect to several conjugate products (such as $\sum xy^{p^i}$) whenever suitable power sums of the dual basis elements do not vanish. The manner in which the condition on the basis separates from the condition on the code and controls self-orthogonality is an illustration of the strong structure of images of codes. In our most general results, self-orthogonality of images of {\it scalable} codes (scalar multiple of a codeword is a codeword; sum of two different codewords need not be a codeword) is studied with respect to a given biadditive form in vector spaces over finite fields. The structure of general biadditive forms over finite fields is exploited in deriving the necessary and sufficient conditions for self-orthogonality. 

The important special case of the canonical inner product ($\sum xy$) in studied in detail. For this case, the following interesting conclusions can be readily shown using our results: (1) Only self-orthogonal codes result in self-orthogonal images in characteristic-2 fields under the canonical inner product. Surprisingly, this result is not true for images over odd-characteristic fields with respect to the canonical inner product. (2) Self-orthogonality of the code is by itself not sufficient to make an image self-orthogonal with respect to the canonical inner product. For many bases of imaging, the code will have to be self-orthogonal with respect to other inner products.

Using our results, we have also studied the relationship between the self-orthogonality of the trace and the image of a code. Since the image of a code is a concatenation of codewords from the trace of the code, the trace of the code plays an important role in determining the orthogonality properties of the image \cite{sakakibara,andrew}. Self-orthogonality of the trace can be determined as a corollary to many of our results concerning images. In particular, we have shown that the trace is self-orthogonal if and only if all images are self-orthogonal with only a single exception of images of codes from GF(4) to GF(2). For the case of quadratic extensions (GF($q^2$) over GF($q$)) and Hermitian inner products, we provide complete analysis that results in a simple criteria to check if an image can be self-orthogonal without the trace being self-orthogonal.

An important application for self-orthogonal codes is in the construction of quantum codes \cite{calderbank}. We expand on the codes provided in \cite{andrew} and provide constructions for a larger set of quantum codes from self-orthogonal GF(4)-images of codes over GF($4^m$). As shown in \cite{calderbank}, quantum error correcting codes can be obtained from linear codes over GF(4) which are self-orthogonal w.r.t the Hermitian inner product $\sum xy^2$. We state the theorem found in \cite{calderbank} for completeness:
\begin{thm}[Calderbank et al \cite{calderbank}]\label{thm:QuantCal}
Suppose $\SC$ is a $(n,k)$ linear code over GF(4) self-orthogonal w.r.t the Hermitian inner product and $d$ is the minimum weight of $\SC^\perp\backslash\SC$. Then, an $[[n, n-2k, d]]$ quantum code can be obtained from $\SC$.
\end{thm}
Hence, an $(n,k,d)$ code over GF$(4^m)$ with 4-ary images self-orthogonal w.r.t the Hermitian inner product leads to an $[[mn,mn-2mk,d']]$ quantum code, where $d'$ is the minimum distance of $\SC^\perp\backslash\SC$. Additionally, $d'\geq d^\perp$, where $d^\perp$ is the minimum distance of $\SC^\perp$.

In \cite{andrew}, cyclic codes over GF$(4^m)$ whose 4-ary traces are self-orthogonal w.r.t the Hermitian inner product have been considered and their images have been used to obtain a class of quantum codes. Theorems \ref{thm:IffImHerm} and \ref{thm:IffTrHerm} below show that, in general, requiring $\tr$ to be self-orthogonal is stronger that requiring $\im$ to be self-orthogonal. We give examples of some RS codes whose 4-ary images are self-orthogonal w.r.t the Hermitian inner product but not the trace thus getting a class of codes larger than that given in \cite{andrew}. This also leads to codes having larger minimum distance for the same codelength than those given in \cite{andrew}.

To the best of our knowledge, all of our above results for self-orthogonality of images and traces of scalable codes with respect to biadditive forms appear to be new. Many results for the special case of the canonical inner product do not appear to be well-known either. A prior work on self-orthogonality of images is \cite{retter1}, where conditions for self-orthogonality of binary images of single-frequency cyclic codes with respect to the canonical inner product have been derived; the conditions in \cite{retter1} are specific to single-frequency cyclic codes over characteristic-2 fields and binary imaging. As stated above, we have studied a much more generalized version of the self-orthogonality problem for images and traces, and derived several interesting and novel results for more general codes and biadditive forms. An illustration of the usefulness of our results is the direct application to the construction of additive quantum error-correcting codes, which require self-orthogonal codes over GF(4) with respect to the Hermitian inner product.

The rest of the paper is organized as follows. We introduce notation and some basic definitions in Section \ref{sec:defn}. Our main results are presented in the form of two theorems in Section \ref{sec:biadditive}. Numerous special cases and interesting results are derived and studied in Section \ref{sec:cases}. The simple case of quadratic extension (GF($q^2$) over GF($q$)) is explored in detail in Section \ref{sec:Quad}. Several examples of self-orthogonal images and construction of new quantum codes is presented in Section \ref{sec:ExQuant}. We conclude in Section \ref{sec:conc} with some discussion of results and remarks.
\section{Definitions and Notation}\label{sec:defn}
We begin by introducing our notation and stating a few relevant preliminary results. See \cite{sloane} as a reference for further details. Let $p$ be a prime number and $q$ a power of $p$ -  i.e., $q=p^r$ for some $r>0$. Let $\q$ denote the finite field with $q$ elements. The finite field $\qm$ is a field extension of degree $m$ of the field $\q$. The trace map $\textrm{Tr}:\qm\to\q$ is defined as Tr$(a) = a+a^q+\ldots+a^{q^{m-1}}$ for $a\in\qm$. Let $\SB=\{\beta_1,\beta_2,\ldots,\beta_m\}$ be a basis of $\qm$ when seen as a vector space over $\q$. Then there exists a unique basis $\SB'=\{\beta'_1,\beta'_2,\ldots,\beta'_m\}$ such that Tr$(\beta_i\beta'_j)=\delta_{ij}$ for $1\leq i, j\leq m$. $\SB'$ is said to be the \emph{dual basis} of $\SB$ and vice versa. $\SB$ is said to be a \emph{self-dual basis} if $\SB'=\SB$. Clearly, $a = $Tr$(\beta'_1 a)\beta_1+$Tr$(\beta'_2 a)\beta_2+\ldots+$Tr$(\beta'_m a)\beta_m$  for all $a\in\qm$. Hence, $(\textrm{Tr}(\beta'_1a),\textrm{Tr}(\beta'_2 a),\ldots,\textrm{Tr}(\beta'_m a))$ are the coordinates of $a\in\qm$ with respect to (w.r.t) the basis $\SB$. 

A \emph{code} $\SC$ over $\qm$ of length $n$ is a subset of $\qm^n$. A \emph{scalable code} is a code $\SC$ such that 
$x\in \SC\Rightarrow \alpha x\in \SC\; \forall \alpha\in \qm$. In other words, a scalable code of length $n$ over $\qm$ is a subset of $\qm^n$ consisting of straight lines through the origin. A \emph{linear code} $\SC$ is a subspace of $\qm^n$ and hence is scalable.

Let $\SB$ and $\SB'$ be as defined above. Define $\textrm{Im}_\SB:\qm^n\to\q^{nm}$ and $\textrm{Tr}:\qm^n\to\q^{n}$ by 
\begin{eqnarray*}
\textrm{Im}_\SB((\alpha_1,\alpha_2,\ldots,\alpha_n)) &=& (\textrm{Tr}(\beta'_1\alpha_1),\ldots,\textrm{Tr}(\beta'_1\alpha_n),\ldots,\textrm{Tr}(\beta'_m\alpha_1),\ldots,\textrm{Tr}(\beta'_m\alpha_n))\\
\textrm{Tr}((\alpha_1,\alpha_2,\ldots,\alpha_n)) &=& (\textrm{Tr}(\alpha_1),\textrm{Tr}(\alpha_2),\ldots,\textrm{Tr}(\alpha_n)).
\end{eqnarray*}
In other words, Im$_\SB$ replaces every coordinate of a vector in $\qm^n$ with its coordinates w.r.t the basis $\SB$ and arranges these coordinates in a specific order and Tr replaces every coordinate of a vector in $\qm^n$ with its trace. $\im$ is called the \emph{Image of }$\SC$ w.r.t the basis $\SB$ and $\tr$ is called the \emph{Trace of }$\SC$. Clearly, $\im$ and $\tr$ are codes over $\q$ of lengths $nm$ and $n$ respectively. Additionally, these codes are scalable (linear) if $\SC$ is scalable (linear). Notice that if we set $\SB'=\{1\}$ (though not a basis) we will get $\tr$ as the {\it image}.

A function $f:\qm^n\times\qm^n\to\qm$ is said to be a \emph{biadditive form} if $f(x+y,z)=f(x,z)+f(y,z)$ and $f(z,x+y)=f(z,x)+f(z,y)$ for all $x,y,z\in$ $\qm^n$. When studying self-orthogonality of traces and images of codes over $\qm$, it is useful to consider two other related biadditive forms. The first form is the natural restriction $f:\q^n\times\q^n\to \qm$. The restricted form is easily seen to be biadditive. The second induced biadditive form $\tilde{f}:\q^{nm}\times\q^{nm}\to\qm$ is defined as 
$$\tilde{f}(x,y)=\sum_{i=0}^{m-1}f((x_{in+1},x_{in+2},\ldots,x_{in+n}),(y_{in+1},y_{in+2},\ldots,y_{in+n})),$$ 
where $x=(x_1,x_2,\ldots,x_{nm}), y=(y_1,y_2,\ldots,y_{nm})$ are vectors in $\q^{mn}$. We say that a code $\SC$ over $\qm$ is \emph{self-orthogonal} w.r.t a biadditive form $f$ if $f(x,y)=0$ for all $x,y\in \SC$. In this work, we consider the problem of determining when  $\im$ and $\tr$ are self-orthogonal w.r.t the induced and restricted biadditive forms $\T{f}$ and $f$, respectively, when $\SC$ is a scalable code.

Two particular cases of biadditive forms are important: if $f$ is defined as $f(x,y) = \sum_{i=1}^n x_iy_i$ then $f$ is called the \emph{canonical inner product} and if $f$ is defined as $f(x,y) = \sum_{i=1}^n x_iy_i^{q^kp^l}$, where $0\leq k\leq m-1$ and $0\leq l\leq r-1$, then it is called a \emph{Hermitian-type product} and is denoted by $f_{kl}$. We note that the induced and restricted forms obtained from the canonical inner product are also canonical inner products. Additionally, the Hermitian-type product defined by $\T{h}_l((x_1,\ldots,x_{mn}),(y_1,\ldots,y_{mn})) = \sum_{i=1}^{mn} x_iy_i^{p^l}$ is the form induced by $f_{kl}$ and the Hermitian-type product defined by $h_l((x_1,\ldots,x_n),(y_1,\ldots,y_n)) = \sum_{i=1}^n x_iy_i^{p^l}$ is the form obtained by restricting the domain of $f$. We consider these special cases and derive results specific to them.

\section{Self-orthogonality w.r.t Biadditive Forms}\label{sec:biadditive}
In this section, we consider self-orthogonality of images and trace of a scalable code w.r.t biadditive forms. We derive the necessary and sufficient condition for self-orthogonality of images and trace and prove that self-orthogonality of image for all bases is equivalent to self-orthogonality of trace. We need two lemmas. The first one concerns the structure of general biadditive forms over finite fields and the forms induced by them.

\begin{lem}\label{lem:StrucBiad}
Let $q=p^r$, where $p$ is a prime, and $f:\qm^n\times\qm^n\to\qm$ be a biadditive form and $\tilde{f}:\q^{nm}\times\q^{nm}\to\qm$ be the biadditive form induced by $f$. Then $$f((x_1,\ldots,x_n),(y_1,\ldots,y_n))=\sum_{1\leq i,j\leq n} \sum_{0\leq k,l\leq rm-1}a_{ijkl}x_i^{p^k}y_j^{p^l},$$ $$\T{f}((x_1,\ldots,x_{nm}),(y_1,\ldots,y_{nm}))=\sum_{1\leq i,j\leq n}\sum_{0\leq k,l\leq r-1}\sum_{s=0}^{m-1}b_{ijkl}x_{sn+i}^{p^k}y_{sn+j}^{p^l},$$ where $a_{ijkl}\in\qm$ and $b_{ijkl}=\sum_{0\leq u,v\leq m-1}a_{ij(k+ur)(l+vr)}$.
\end{lem} 
\begin{proof}
Since $f$ is biadditive, $f(ax,by)=ab(x,y)$ for all $a,b\in$ $\p$ and $x,y\in\qm^n$. Let $\{\beta_1,\ldots,\beta_{rm}\}$ be a basis of $\qm$ over $\p$ and $\{\beta'_1,\ldots,\beta'_{rm}\}$ be its dual basis. Let $\{e_1,\ldots,e_n\}$ be the standard basis of $\qm^n$ over $\qm$. Then $(x_1,\ldots,x_n) = \sum_{i=1}^n x_ie_i$ and $a = \sum_{s=1}^{rm} \textrm{Tr}(\beta'_s a)\beta_s$ (here the trace map is from $\qm$ to GF($p$)) for all $(x_1,\ldots,x_n)\in$ $\qm^n$ and $a\in$ $\qm $. Hence, 
\begin{eqnarray*}
f((x_1,\ldots,x_n),(y_1,\ldots,y_n))& = &f(\sum_{i=1}^n\sum_{s=1}^{rm}\textrm{Tr}(\beta'_s x_i)\beta_se_i, \sum_{j=1}^n\sum_{t=1}^{rm}\textrm{Tr}(\beta'_t y_j)\beta_te_j)\\
& = & \sum_{1\leq i,j\leq n}\sum_{1\leq s,t\leq rm} \textrm{Tr}(\beta'_s x_i)\textrm{Tr}(\beta'_t y_j) f(\beta_s e_i, \beta_t e_j).
\end{eqnarray*}
Since Tr$(\beta'_sx_i)= \beta'_sx_i + (\beta'_sx_i)^p+\ldots+(\beta'_sx_i)^{p^{r m-1}}$ and Tr$(\beta'_ty_j)= \beta'_ty_j + (\beta'_ty_j)^p+\ldots+(\beta'_ty_j)^{p^{rm-1}}$, we have 
\begin{eqnarray*}
f((x_1,\ldots,x_n),(y_1,\ldots,y_n))&=&\sum_{1\leq i,j\leq n}\sum_{1\leq s,t\leq rm} \sum_{0\leq k,l\leq rm-1}(\beta'_s x_i)^{p^k}(\beta'_t y_j)^{p^l}f(\beta_s e_i, \beta_t e_j)\\
& = &\sum_{1\leq i,j\leq n}\sum_{0\leq k,l\leq rm-1}a_{ijkl}x_i^{p^k}y_j^{p^l},
\end{eqnarray*}
where $a_{ijkl} = \sum_{1\leq s,t\leq rm} \beta_s^{'p^k}\beta_t ^{'p^l}f(\beta_s e_i, \beta_t e_j)$. By definition,  $$\T{f}((x_1,\ldots,x_{nm}),(y_1,\ldots,y_{nm}))=\sum_{s=0}^{m-1}\sum_{1\leq i,j\leq n}\sum_{0\leq k,l\leq rm-1}a_{ijkl}x_{sn+i}^{p^k}y_{sn+j}^{p^l}.$$ Since the coordinates satisfy $X^q=X^{p^r}=X$, we have $$\T{f}((x_1,\ldots,x_{nm}),(y_1,\ldots,y_{nm}))=\sum_{s=0}^{m-1}\sum_{1\leq i,j\leq n}\sum_{0\leq k,l\leq r-1}b_{ijkl}x_{sn+i}^{p^k}y_{sn+j}^{p^l},$$ where $b_{ijkl}=\sum_{0\leq u,v\leq m-1}a_{ij(k+ur)(l+vr)}$.
\end{proof}
The second lemma is a property of the trace map.

\begin{lem}\label{lem:TrPoly}
Let $\textrm{Tr}:\qm\to\q$ be the trace map and $a_0,\ldots,a_{q-1}$ be elements of $\qm$. Then $\textrm{Tr}(a_0+\lambda a_1+ \lambda^2 a_2+\ldots +\lambda^{q-1}a_{q-1}) = 0$ for all $\lambda \in$ $\qm$ if and only if $\textrm{Tr}(a_0),a_1,\ldots,a_{q-1} $ are all zero.
\end{lem}
\begin{proof}
\begin{eqnarray*}
&&\textrm{Tr}(a_0+\lambda a_1+ \lambda^2 a_2+\ldots +\lambda^{q-1}a_{q-1}) = \\
&&a_0+\lambda a_1+ \lambda^2 a_2+\ldots +\lambda^{q-1}a_{q-1}+\\
&&a_0^q+\lambda^q a_1^q+ \lambda^{2q} a_2^q+\ldots +\lambda^{q(q-1)}a_{q-1}^q+\ldots\\
&&a_0^{q^{m-1}}+\lambda^{q^{m-1}} a_1^{q^{m-1}}+ \lambda^{2q^{m-1}} a_2^{q^{m-1}}+\ldots +\lambda^{q^{m-1}(q-1)}a_{q-1}^{q^{m-1}}
\end{eqnarray*}
Hence, we have $q^m$ zeros for a polynomial of degree at most $q^{m-1}(q-1)$ with coefficients in $\qm$. This is possible if and only if all the coefficients are zero. Equating the constant term to zero, we get $\textrm{Tr}(a_0)=0$. Equating the coefficients of $\lambda$, $\lambda^2$,$\cdots$,$\lambda^{q-1}$ to zero, we get $a_1,\ldots,a_{q-1} $ are all zero. 
\end{proof}
\subsection{Self-orthogonality of images and traces of codes}
We now state our main result concerning the self-orthogonality of images of codes in the following theorem.
\begin{thm}[Self-orthogonality of $\im$]\label{thm:IffImBiad}
Let $\SC$ be a scalable code over $\qm$ of length $n$. Let $q=p^r$, where $p$ is a prime number. Let $\SB$ be a basis of $\qm$ over $\q$ and $\SB'=\{\beta_1,\ldots,\beta_m\}$ be its dual basis. Let $f:\qm^n\times\qm^n\to\qm$ be given by $$f((x_1,\ldots,x_n),(y_1,\ldots,y_n))=\sum_{1\leq i,j\leq n}\sum_{0\leq k,l\leq rm-1}a_{ijkl}x_i^{p^k}y_j^{p^l}$$ for some $a_{ijkl}\in$ $\qm$. Let $\T{f}:\q^{mn}\times\q^{mn}\to\qm$ be the biadditive form induced by $f$. Let $b_{ijkl}=\sum_{0\leq u,v\leq m-1}a_{ij(k+ur)(l+vr)}$. Then $\im$ is self-orthogonal w.r.t $\T{f}$ if and only if $$(\sum_{1\leq i,j\leq n}b_{ijkl} x_iy_j^{p^{l-k}q^w})(\sum_{s=1}^m\beta_s^{1+p^{l-k}q^w}) = 0$$ for all $(x_1,\ldots,x_n),(y_1,\ldots,y_n)\in \SC, 0\leq k,l\leq r-1$ and $0\leq w\leq m-1$
\end{thm}
\begin{proof}
From Lemma \ref{lem:StrucBiad}, $\im$ is self-orthogonal w.r.t $\T{f}$ if and only if
$$\sum_{1\leq i,j\leq n}\sum_{0\leq k,l\leq r-1}\sum_{s=1}^m b_{ijkl} \textrm{Tr}(\beta_s x_i)^{p^k}\textrm{Tr}(\beta_s y_j)^{p^l} = 0 \quad \forall x=(x_1,\ldots,x_n),y=(y_1,\ldots,y_n)\in \SC.$$
Since $\SC$ is a scalable code, the above condition is equivalent to 
$$\sum_{1\leq i,j\leq n}\sum_{0\leq k,l\leq r-1}\sum_{s=1}^m b_{ijkl} \textrm{Tr}(\beta_s\lambda_1x_i)^{p^k}\textrm{Tr}(\beta_s\lambda_2 y_j)^{p^l} = 0 \quad \forall x,y \in \SC, \lambda_1,\lambda_2\in \qm.$$
Let $\{c_{ijklt}\}_{1\leq t\leq m}$ be the coordinates of $b_{ijkl}$ w.r.t some basis $\{\gamma_1,\gamma_2,\ldots,\gamma_m\}$ of $\qm$ over $\q$. Writing $b_{ijkl}$ as $\sum_{t=1}^m c_{ijklt}\gamma_t$ we get that the above condition is equivalent to 
$$\sum_{t=1}^m\Bigg\{\sum_{\substack{1\leq i,j\leq n\\0\leq k,l\leq r-1\\ 1\leq s\leq m}}c_{ijklt} \textrm{Tr}(\beta_s\lambda_1x_i)^{p^k}\textrm{Tr}(\beta_s\lambda_2 y_j)^{p^l}\Bigg\}\gamma_t = 0 \quad \forall x,y\in \SC, \lambda_1,\lambda_2\in \qm.$$
Each term in the parenthesis is an element of $\q$ and $\{\gamma_1,\gamma_2,\ldots,\gamma_m\}$ is a basis of $\qm$ over $\q$. Hence, the above sum vanishes if and only if each term in the parenthesis vanishes. In other words, the above condition is equivalent to
$$\sum_{\substack{1\leq i,j\leq n\\0\leq k,l\leq r-1\\ 1\leq s\leq m}}c_{ijklt} \textrm{Tr}(\beta_s\lambda_1x_i)^{p^k}\textrm{Tr}(\beta_s\lambda_2 y_j)^{p^l} = 0 \quad \forall x,y\in \SC, 1\leq t\leq m, \lambda_1,\lambda_2\in \qm.$$
Using the definition of Tr and the fact that it is a linear functional from $\qm$ to $\q$ we have 
$$\sum_{\substack{1\leq i,j\leq n\\0\leq k,l\leq r-1\\ 1\leq s\leq m}}c_{ijklt} \textrm{Tr}(\beta_s\lambda_1x_i)^{p^k}\textrm{Tr}(\beta_s\lambda_2 y_j)^{p^l} = \textrm{Tr}(\sum_{\substack{1\leq i,j\leq n\\0\leq k,l\leq r-1\\1\leq s\leq m\\0\leq w\leq m-1}}c_{ijklt} (\beta_s\lambda_1x_i)^{p^k}(\beta_s\lambda_2 y_j)^{p^{l+wr}}).$$
Hence, we need trace of a polynomial in $\lambda_1$ of degree at most $p^{r-1}$ to be identically zero for all $\lambda_1\in  \qm$. By Lemma \ref{lem:TrPoly}, this is possible if and only if each coefficient of the polynomial is zero. Hence, $\im$ is self-orthogonal w.r.t $\T{f}$ if and only if
$$ \sum_{\substack{1\leq i,j\leq n\\0\leq l\leq r-1\\1\leq s\leq m\\0\leq w\leq m-1}}c_{ijklt} (\beta_sx_i)^{p^k}(\beta_s\lambda_2 y_j)^{p^{l+wr}} = 0 \quad \forall x,y\in \SC, 0\leq k\leq r-1,1\leq t\leq m,\lambda_2\in \qm.$$
Since $b_{ijkl} = \sum_{t=1}^m c_{ijklt}\gamma_t$, the above condition is equivalent to 
$$ \sum_{\substack{1\leq i,j\leq n\\0\leq l\leq r-1\\1\leq s\leq m\\0\leq w\leq m-1}}b_{ijkl} (\beta_sx_i)^{p^k}(\beta_s\lambda_2 y_j)^{p^{l+wr}} = 0 \quad \forall x,y\in \SC, 0\leq k\leq r-1, \lambda_2\in \qm.$$
Hence, we need $p^{rm}$ zeros for a polynomial in $\lambda_2$ of degree at most $p^{rm-1}$ with coefficients in $\textrm{GF}(p^{rm})$. This is possible if and only if all the coefficients are zero - i.e., if and only if 
$$\sum_{1\leq i,j\leq n}\sum_{s=1}^m b_{ijkl} (\beta_sx_i)^{p^k}(\beta_sy_j)^{p^{l+wr}} = 0 \quad \forall x,y\in \SC, 0\leq k,l\leq r-1, 0\leq w\leq m-1.$$
Hence, $\im$ is self-orthogonal w.r.t $f$ if and only if $$(\sum_{1\leq i,j\leq n}b_{ijkl} x_i^{p^k}y_j^{p^lq^w})(\sum_{s=1}^{m}\beta_s^{p^k+p^lq^w}) = 0 \quad \forall x,y\in \SC, 0\leq k,l\leq r-1\textrm{ and }0\leq w\leq m-1.$$
Since, every element in $\qm$ has a $p$th root and $\qm$ is of characteristic $p$, $\im$ is self-orthogonal w.r.t $f$ if and only if $$(\sum_{1\leq i,j\leq n}b_{ijkl} x_iy_j^{p^{l-k}q^w})(\sum_{s=1}^{m}\beta_s^{1+p^{l-k}q^w}) = 0$$ for all $(x_1,\ldots,x_n),(y_1,\ldots,y_n)\in \SC, 0\leq k,l\leq r-1$ and $0\leq w\leq m-1$
\end{proof}

Notice that in the above proof, the fact that $\SB'$ is a basis is never used. Hence, setting $\SB'=\{1\}$ we get our most general result concerning self-orthogonality of traces of codes.
\begin{thm}[Self-orthogonality of $\tr$]\label{thm:IffTrBiad}
Let $\SC$ be a code over $\qm$. Let $q=p^r$, where $p$ is a prime number. Let $f:\qm^n\times\qm^n\to\qm$ be given by $$f((x_1,\ldots,x_n),(y_1,\ldots,y_n))=\sum_{1\leq i,j\leq n}\sum_{0\leq k,l\leq rm-1}a_{ijkl}x_i^{p^k}y_j^{p^l}$$ for some $a_{ijkl}\in$ $\qm)$ and $b_{ijkl}=\sum_{0\leq u,v\leq m-1}a_{ij(k+ur)(l+vr)}$. Then $\tr$ is self-orthogonal w.r.t $f$ if and only if $$\sum_{1\leq i,j\leq n}b_{ijkl} x_iy_j^{p^{l-k}q^w} = 0$$ for all $(x_1,\ldots,x_n),(y_1,\ldots,y_n)\in \SC, 0\leq k,l\leq r-1$ and $0\leq w\leq m-1$
\end{thm}

The above two results say the following: given a basis for $\textrm{GF}(p^{rm})$ over $\textrm{GF}(p^r)$ and a biadditive form $f$, we have $r^2m$ related conjugate biadditive forms and $r^2m$ power sums of the dual basis elements corresponding to each value of $k$, $l$ and $w$. $\im$ is self-orthogonal if and only if $\SC$ is self-orthogonal w.r.t all those biadditive forms for which the corresponding power sum of the dual basis elements is non-zero and $\tr$ is self-orthogonal if and only if $\SC$ is self-orthogonal w.r.t all the $r^2m$ biadditive forms. We note that for a fixed $k$ and $l$, all the $m$ power sums $\sum_{s=1}^{m}\beta_s^{1+p^{l-k}q^w}, 0\leq w\leq m-1$ cannot be zero. ( $\sum_{w=0}^{m-1}\sum_{s=1}^{m}\beta_s^{1+p^{l-k}q^w}  = \sum_{s=1}^m \rm{Tr}(\beta_s)^{p^{l-k}}\beta_s \neq 0$, since $\SB'$ is a basis for $\qm$ over  $\q$ and Tr  is a non-zero linear functional from $\qm$ to $\q$.) Hence, $\im$ being self-orthogonal forces $\SC$ to be self-orthogonal w.r.t at least $r^2$ biadditive forms. We note that some or all of these forms might be identically zero depending on $f$. For example, let $q$ be even and $f$ be given by $f(x,y) = \sum_{i=1}^n x_iy_i+x_iy_i^q$. Then $\T{f}$ is the zero map.
\subsection{Self-orthogonality of images w.r.t. all bases}
We now prove the equivalence of self-orthogonality of image for all bases and self-orthogonality of trace. By definition, each codeword of $\im$ is got by concatenating certain codewords of $\tr$. As observed in \cite{andrew}, if $\tr$ is self-orthogonal w.r.t $f$ then $\im$ is self-orthogonal w.r.t $\T{f}$ for every basis $\SB$. The following two results show that the converse is also true except for the case $q=m=2$. We give an example to show that the converse need not hold when $q=m=2$. Later, we examine why this happens.

\begin{thm}\label{thm:q>2ImTrBiad}
Let $\SC$ be a scalable code of length $n$ over $\qm$. Let $f:\qm^n\times\qm^n\to\qm$ be a biadditive form and $\T{f}:\q^{mn}\times\q^{mn}\to\qm$ be the biadditive form induced by $f$. Suppose $q>2$ and $\im$ is self-orthogonal w.r.t $\T{f}$ for three bases $\SB_1,\SB_2,\SB_3$ of $\qm$ over $\q$ such that $\SB_1'=\{\beta_1,\ldots,\beta_m\},  \SB_2'=\{\beta_1+\alpha\beta_2,\beta_2,\ldots,\beta_m\}$ and $\SB_3'=\{\beta_1+\gamma\beta_2,\beta_2,\ldots,\beta_m\}$, where $\alpha$ and $\gamma$ are distinct non-zero elements of $\q$. Then $\tr$ is self-orthogonal w.r.t $f$ and $\im$ is self-orthogonal w.r.t $\T{f}$ for all bases $\SB$.
\end{thm}
\begin{proof}
From Theorems \ref{thm:IffImBiad} and \ref{thm:IffTrBiad}, to prove that $\tr$ is self-orthogonal w.r.t. $f$ it is enough to show that for all $0\leq k,l\leq r-1$ and $0\leq w\leq m-1$ one of the following equations is false:
\begin{eqnarray}
\sum_{s=1}^m\beta_s^{1+p^{l-k}q^w}&=&0 \label{eqn:4.1}\\
(\beta_1+\alpha\beta_2)^{1+p^{l-k}q^w}+\sum_{s=2}^m\beta_s^{1+p^{l-k}q^w} &=&0\label{eqn:4.2}\\ (\beta_1+\gamma\beta_2)^{1+p^{l-k}q^w}+\sum_{s=2}^m\beta_s^{1+p^{l-k}q^w} &=&0.\label{eqn:4.3}
\end{eqnarray}
Suppose all the above three equations are true for some $k$, $l$ and $w$. Using the fact that $\qm$ is of characteristic $p$ and comparing (\ref{eqn:4.1}) and (\ref{eqn:4.2}) and (\ref{eqn:4.1}) and (\ref{eqn:4.3}) we have, 
\begin{eqnarray}
\alpha\beta_2\beta_1^{p^{l-k}q^w}+\alpha^{p^{l-k}q^w}\beta_1\beta_2^{p^{l-k}q^w}+(\alpha\beta_2)^{1+p^{l-k}q^w} &=&0.\label{eqn:4.4}\\
\gamma\beta_2\beta_1^{p^{l-k}q^w}+\gamma^{p^{l-k}q^w}\beta_1\beta_2^{p^{l-k}q^w}+(\gamma\beta_2)^{1+p^{l-k}q^w} &=&0.\label{eqn:4.5}
\end{eqnarray}
Multiplying (\ref{eqn:4.4}) by $\gamma$ and (\ref{eqn:4.5}) by $\alpha$, subtracting one from the other and dividing the resulting equation by $\beta_2^{p^{l-k}q^w}$ we get $$(\gamma\alpha^{p^{l-k}q^w}-\alpha\gamma^{p^{l-k}q^w})\beta_1+(\gamma\alpha^{1+p^{l-k}q^w}-\alpha\gamma^{1+p^{l-k}q^w})\beta_2 = 0.$$
Since $\beta_1$ and $\beta_2$ are linearly independent over $\q$ we have $\gamma\alpha^{p^{l-k}q^w}=\alpha\gamma^{p^{l-k}q^w}$ and $\gamma\alpha^{1+p^{l-k}q^w}=\alpha\gamma^{1+p^{l-k}q^w}$. Since $\alpha$ and $\gamma$ are distinct and non-zero these equations lead to a contradiction. It follows that $\tr$ is self-orthogonal w.r.t $f$, hence $\im$ is self-orthogonal w.r.t $\T{f}$ for all bases $\SB$.
\end{proof}
Notice that the condition $q>2$ is vital for the above theorem as two distinct nonzero elements are assumed to be available in the field. We next prove a similar result for the case $m>2$.

\begin{thm}\label{thm:m>2ImTrBiad}
Let $\SC$ be a scalable code of length $n$ over $\qm$. Let $f:\qm^n\times\qm^n\to\qm$ be a biadditive form and $\T{f}:\q^{mn}\times\q^{mn}\to\qm$ be the biadditive form induced by $f$. Suppose $m>2$ and $\im$ is self-orthogonal w.r.t $\T{f}$ for five bases $\SB_1,\SB_2,\SB_3,\SB_4,\SB_5$ of $\qm$ over $\q$ such that $\SB_1'=\{\beta_1,\ldots,\beta_m\}, \SB_2'=\{\beta_1+\alpha\beta_2,\beta_2,\ldots,\beta_m\}$, $\SB_3'=\{\beta_1+\gamma\beta_3,\beta_2,\ldots,\beta_m\}$, $\SB_4'=\{\beta_1,\beta_2+\delta\beta_3,\beta_3,\ldots,\beta_m\},$ and  $\SB_5'=\{\beta_1+\alpha\beta_2+\gamma\beta_3,\beta_2,\ldots,\beta_m\} $, where $\alpha,\gamma,\delta$ are  non-zero not necessarily distinct elements of $\q$. Then $\tr$ is self-orthogonal w.r.t $f$ and $\im$ is self-orthogonal w.r.t $\T{f}$ for all bases.
\end{thm}
\begin{proof}
From Theorems \ref{thm:IffImBiad} and \ref{thm:IffTrBiad}, to prove that $\tr$ is self-orthogonal w.r.t. $f$ it is enough to show that for all $0\leq k,l\leq r-1$ and $0\leq w\leq m-1$ one of the following equations is false:
\begin{eqnarray}
\sum_{s=1}^m\beta_s^{1+p^{l-k}q^w} &=& 0 \label{eqn:4.6}\\
(\beta_1+\alpha\beta_2)^{1+p^{l-k}q^w}+\sum_{s=2}^m\beta_s^{1+p^{l-k}q^w} &=& 0 \label{eqn:4.7}\\ (\beta_1+\gamma\beta_3)^{1+p^{l-k}q^w}+\sum_{s=2}^m\beta_s^{1+p^{l-k}q^w} &=& 0 \label{eqn:4.8}\\
\beta_1^{1+p^{l-k}q^w}+(\beta_2+\delta\beta_3)^{1+p^{l-k}q^w}+\sum_{s=3}^m\beta_s^{1+p^{l-k}q^w} &=& 0 \label{eqn:4.9}\\
(\beta_1+\alpha\beta_2+\gamma\beta_3)^{1+p^{l-k}q^w}+\sum_{s=2}^m\beta_s^{1+p^{l-k}q^w} &=& 0. \label{eqn:4.10}
\end{eqnarray}
Suppose all the above five equations are true for some $k$, $l$ and $w$. Using the fact that $\qm$ is of characteristic $p$ and comparing (\ref{eqn:4.6}) with each of (\ref{eqn:4.7}), (\ref{eqn:4.8}), (\ref{eqn:4.9}) and (\ref{eqn:4.10}) we have,
\begin{eqnarray}
\alpha \beta_2 \beta_1^{p^{l-k}q^w} +\alpha^{p^{l-k}q^w} \beta_1 \beta_2^{p^{l-k}q^w} + (\alpha \beta_2)^{1+p^{l-k}q^w} & = & 0,\label{eqn:4.11}\qquad\\
\gamma \beta_3 \beta_1^{p^{l-k}q^w} +\gamma^{p^{l-k}q^w} \beta_1 \beta_3^{p^{l-k}q^w} + (\gamma \beta_3)^{1+p^{l-k}q^w} & = & 0,\label{eqn:4.12}\\
\delta \beta_3 \beta_2^{p^{l-k}q^w} +\delta^{p^{l-k}q^w} \beta_2 \beta_3^{p^{l-k}q^w} + (\delta \beta_3)^{1+p^{l-k}q^w} & = & 0,\label{eqn:4.13}\\
\alpha^{p^{l-k}q^w} \beta_1 \beta_2^{p^{l-k}q^w} +\gamma^{p^{l-k}q^w} \beta_1 \beta_3^{p^{l-k}q^w} + \alpha \beta_2 \beta_1^{p^{l-k}q^w} + (\alpha \beta_2)^{1+p^{l-k}q^w} + & & \nonumber\\
\alpha\gamma^{p^{l-k}q^w} \beta_2 \beta_3^{p^{l-k}q^w} + \gamma \beta_3 \beta_1^{p^{l-k}q^w} + \gamma\alpha^{p^{l-k}q^w} \beta_3 \beta_2^{p^{l-k}q^w} + (\gamma \beta_3)^{1+p^{l-k}q^w} & = &  0.\label{eqn:4.14}
\end{eqnarray}
From (\ref{eqn:4.11}), (\ref{eqn:4.12}) and (\ref{eqn:4.14}) above we have 
\begin{equation}
\alpha\gamma^{p^{l-k}q^w} \beta_2 \beta_3^{p^{l-k}q^w}+\gamma\alpha^{p^{l-k}q^w} \beta_3 \beta_2^{p^{l-k}q^w} = 0.\label{eqn:4.15}
\end{equation}
Multiplying (\ref{eqn:4.15}) by $\delta$ and (\ref{eqn:4.13}) by $\gamma\alpha^{p^{l-k}q^w}$, subtracting one from the other and dividing the resulting equation by $\beta_3^{p^{l-k}q^w}$ we get 
$$(\gamma(\alpha\delta)^{p^{l-k}q^w}-\alpha\delta\gamma^{p^{l-k}q^w})\beta_2+\gamma\alpha^{p^{l-k}q^w}\delta^{1+p^{l-k}q^w}\beta_3 = 0.$$
Since $\beta_2$ and $\beta_3$ are linearly independent over $\q$ we have $\gamma\alpha^{p^{l-k}q^w}\delta^{1+p^{l-k}q^w}=0$ which is a contradiction to the fact that $\alpha,\gamma$ and $\delta$ are non-zero. It follows that $\tr$ is self-orthogonal w.r.t $f$, hence $\im$ is self-orthogonal w.r.t $\T{f}$ for all bases $\SB$.
\end{proof}
Notice that the condition $m>2$ has been used in the above theorem through the implicit assumption that a basis contains at least three elements $\beta_1$, $\beta_2$ and $\beta_3$. We now see that if either $q>2$ or $m>2$, all images being self-orthogonal implies that trace is self-orthogonal. The only remaining case is that of images of codes over the field with $q=2$ and $m=2$, namely GF(4) over GF(2).

When $q=2$ and $m=2$, $\tr$ need not be self-orthogonal even if $\im$ is self-orthogonal for all bases. Consider $\SC=\{ (0,0,0), (1,\omega,\omega^2), (\omega,\omega^2,1),(\omega^2,1,\omega)\}$, where $\omega$ is a primitive element of GF(4). The three bases for GF(4) over GF(2) are $\SB_1=\{1,\omega\}, \SB_2=\{\omega,\omega^2\}, \SB_3= \{1,\omega^2\}$. It is easily seen that 
\begin{eqnarray*}
\textrm{Im}_{\SB_1} (\SC) &=& \{ (0,0,0,0,0,0),(1,0,0,1,1,1),(0,1,1,1,1,0),(1,1,1,0,0,1)\},\\
\textrm{Im}_{\SB_2} (\SC) &=& \{ (0,0,0,0,0,0),(1,1,1,0,0,1),(1,0,0,1,1,1),(0,1,1,1,1,0)\},\\
\textrm{Im}_{\SB_3} (\SC) &=& \{ (0,0,0,0,0,0),(1,0,1,1,0,1),(1,1,0,1,1,0),(0,1,1,0,1,1)\}.
\end{eqnarray*}
Hence, all the three images are self-orthogonal w.r.t the canonical inner product but 
$$\tr=\{(0,0,0),(0,1,1),(1,1,0),(1,0,1)\}$$ 
and it is not self-orthogonal w.r.t the canonical inner product.
\section{Some Special Cases}\label{sec:cases}
In this section, we apply our main results to various specific situations to derive some results of interest.
\subsection{Self-orthogonality w.r.t Hermitian-type products} 
We begin by considering self-orthogonality of images and trace of a scalable code w.r.t Hermitian-type products due to their importance. Let $q= p^r$, where $p$ is a prime number. For $0\leq k\leq m-1$ and $0\leq l\leq r-1$, a Hermitian-type product $f_{kl}:\qm^n\times\qm^n\to\qm$ is defined as $f_{kl}(x,y) = \sum_{i=1}^n x_iy_i^{p^lq^k}$, where $ x=(x_1,\ldots,x_n), y=(y_1,\ldots,y_n)$. Then the map $\T{h}_l:\q^{mn}\times\q^{mn}\to\q$ given by $\T{h}_{l}(x,y) = \sum_{i=1}^{mn} x_iy_i^{p^l}$ is the map induced by $f_{kl}$ and the restricted map $h_l:\q^{n}\times\q^{n}\to\q$ is given by $h_{l}(x,y) = \sum_{i=1}^n x_iy_i^{p^l}$. Notice that the form $f_{00}$ is the canonical inner product $\sum_{i=1}^n x_iy_i$, which results in both the restricted and induced maps being canonical as well.

We now restate our main results for the case of Hermitian-type products in the following two theorems for ease of reference and clarity.
\begin{thm}[Self-orthogonality of $\im$]\label{thm:IffImHerm}
Let $\SC$ be a scalable code of length $n$ over $\qm$, $\SB$ be a basis of $\qm$ over $\q$ and $\SB'=\{\beta_1,\ldots,\beta_m\}$ be the dual basis of $\SB$. Then $\im$ is self-orthogonal w.r.t the Hermitian-type product, $\sum_{i=1}^{mn} x_iy_i^{p^l}$ if and only if $$( \sum_{i=1}^n x_i y_i^{p^lq^k})( \sum_{j=1}^m\beta_j^{1+p^lq^k})=0$$ for all $x=(x_1,\ldots,x_n),y=(y_1,\ldots,y_n)\in \SC$ and $0\leq k\leq m-1$.
\end{thm}
\begin{thm}[Self-orthogonality of $\tr$]\label{thm:IffTrHerm}
Let $\SC$ be a scalable code of length $n$ over $\qm$. Then $\tr$ is self-orthogonal w.r.t the Hermitian-type product, $\sum_{i=1}^{n} x_iy_i^{p^l}$ if and only if $$\sum_{i=1}^n x_i y_i^{p^lq^k}=0$$ for all $x=(x_1,\ldots,x_n),y=(y_1,\ldots,y_n)\in \SC$  and $0\leq k\leq m-1$ - i.e., if and only if $\SC$ is self-orthogonal w.r.t $f_{kl}$ for $0\leq k \leq m-1$.
\end{thm}

The above two main results say the following: given a basis for $\qm$ over $\q$ and the Hermitian-type product $\sum_{i=1}^{mn} x_iy_i^{p^l}$ over $\q$, we have $m$ related Hermitian-type products $\sum_{i=1}^n x_iy_i^{p^lq^k}$ over $\qm$ and $m$ power sums of the elements of the dual basis $\sum_{j=1}^m\beta_j^{1+p^lq^k}$ corresponding to each value of $k=0,1,\ldots,m-1$. $\im$ is self-orthogonal if and only if $\SC$ is self-orthogonal w.r.t all those Hermitian-type products for which the corresponding power sum of the dual basis elements is non-zero and $\tr$ is self-orthogonal if and only if $\SC$ is self-orthogonal w.r.t all the $m$ Hermitian-type products. For a fixed $l$, all the $m$ power sums $\sum_{j=1}^{m}\beta_j^{1+p^lq^k}, 0\leq k\leq m-1$ cannot be zero. Hence, $\im$ being self-orthogonal forces $\SC$ to be self-orthogonal w.r.t at least one Hermitian-type product.
\subsection{Self-orthogonality w.r.t canonical inner product}
We now derive some interesting results for the case of the canonical inner product. Our interest is in finding non-self-orthogonal codes whose images are self-orthogonal w.r.t the canonical inner product. Most of our results are negative in this context.
\subsubsection{GF(4) over GF(2)}
We have seen that images from GF(4) to GF(2) make an important counterexample for the situation where self-orthogonality w.r.t all bases does not imply self-orthogonality of the trace. 
\begin{prop}\label{prop:GF4overGF2}
Let $\SC$ be a scalable code over GF(4). Then the following are equivalent:\\
(i) $\im$ is self-orthogonal w.r.t the canonical inner product for some basis $\SB$.\\
(ii) $\im$ is self-orthogonal w.r.t the canonical inner product for all bases $\SB$.\\
(iii) $\SC$ is self-orthogonal w.r.t the canonical inner product.
\end{prop}
\begin{proof}
The only bases for GF(4) over GF(2) are $\SB_1=\{1,\omega\},\SB_2=\{1,\omega^2\}$, and $\SB_3=\{\omega,\omega^2\}$, where $\omega$ is a primitive element of GF(4).  By simple computation, it is seen that $\beta_1^{1+2^k}+\beta_2^{1+2^k}$ is non-zero for $k=0$ and zero for $k=1$ for the above three bases. It follows from this and Theorem \ref{thm:IffImHerm} that for any basis $\SB$, $\im$ is self-orthogonal w.r.t the canonical inner product if and only if $\SC$ is self-orthogonal w.r.t the canonical inner product. It follows that the proposition is true.

\emph{Alternate Proof (without using our results). }From the definition of the trace map, it is seen that Tr(0)=0, Tr(1)=1, Tr$(\omega)$=1, and Tr$(\omega^2)$=1. Additionally, the trace map is given by Tr$(a) = a+a^2$ and $a^4=a$ for all $a$ in GF(4). Hence, if $x$ and $y$ are two elements of GF(4),
\begin{eqnarray*}
\textrm{Tr}(x)\textrm{Tr}(y)+\textrm{Tr}(\omega^2 x)\textrm{Tr}(\omega^2 y) &=& \textrm{Tr}(\omega^2 xy),\\
\textrm{Tr}(\omega^2 x)\textrm{Tr}(\omega^2 y)+\textrm{Tr}(\omega x)\textrm{Tr}(\omega y) &=& \textrm{Tr}(xy),\\
\textrm{Tr}(\omega x)\textrm{Tr}(\omega y) +\textrm{Tr}(x)\textrm{Tr}(y) &=& \textrm{Tr}(\omega xy).
\end{eqnarray*}
Suppose $\SB=\SB_1$. Then $\SB'=\{\omega^2,1\}$. Hence, $\im$ is self-orthogonal w.r.t the canonical inner product if and only if $\sum_{i=1}^n \textrm{Tr}(a_i)\textrm{Tr}(b_i)+\textrm{Tr}(\omega^2 a_i)\textrm{Tr}(\omega^2 b_i)= 0$ for all $(a_i), (b_i) \in \SC$. This is equivalent to $\textrm{Tr}(\sum_{i=1}^n\omega^2 a_i b_i) =0$ for all $(a_i), (b_i) \in \SC$. This is true if and only if $\sum_{i=1}^n a_ib_i = \omega$ or 0 for all $(a_i), (b_i) \in \SC$. Suppose $\sum_{i=1}^n a_ib_i=\omega$ for some $(a_i),(b_i)\in \SC$. Since $\SC$ is scalable, $(a_i) \in \SC$ implies $(\omega a_i)\in \SC$. In that case, $\sum_{i=1}^n (\omega a_i)b_i= \omega^2$, which is not possible. Hence, $\im$ is self-orthogonal w.r.t the canonical inner product if and only if $\sum_{i=1}^n a_ib_i = 0$ for all $(a_i), (b_i) \in \SC$ - i.e., if and only if $\SC$ is self-orthogonal w.r.t the canonical inner product. Similarly, if $\SB=\SB_2$ and $\SB_3$ respectively, then $\SB'=\{\omega,1\}$ and $\SB_3$ respectively and $\im$ is self-orthogonal w.r.t the canonical inner product if and only if $\SC$ is self-orthogonal w.r.t to the canonical inner product. Hence, $(i)$ is equivalent to $(iii)$. From this it follows that $(ii)$ and $(iii)$ are equivalent and we are done.
\end{proof}
Let us examine the counterexample more closely. From Theorem \ref{thm:IffTrHerm}, $\tr$ is self-orthogonal w.r.t the canonical inner product if and only if $\SC$ is self-orthogonal w.r.t the canonical and the Hermitian inner products given by $\sum x_i y_i$ and $\sum x_i y_i^2$, respectively.  From Proposition \ref{prop:GF4overGF2}, $\im$ is self-orthogonal w.r.t the canonical inner product for all bases $\SB$ if and only if $\SC$ is self-orthogonal w.r.t the canonical inner product. Hence, we see that $\tr$ being self-orthogonal is a more stringent condition than $\im$ being self-orthogonal for all bases. Hence, for $q=m=2$, we can say $\tr$ is self-orthogonal w.r.t the canonical inner product if and only if $\im$ is self-orthogonal w.r.t the canonical inner product for some basis and $\SC$ is self-orthogonal w.r.t the Hermitian inner product.
\subsubsection{GF($2^m$) over GF(2)}
An interesting result for fields of even characteristic is that self-orthogonality of any image w.r.t the canonical inner product implies self-orthogonality of the original code.
\begin{prop}\label{prop:evenfield}
Let $\SC$ be a scalable code over $\qm$ for some even $q$ and $\SB$ be a basis of $\qm$ over $\q$. If $\im$ is self-orthogonal w.r.t the canonical inner product, then so is $\SC$.
\end{prop}
\begin{proof}
Let $\SB'=\{\beta_1,\ldots,\beta_m\}$. From Theorem \ref{thm:IffImHerm}, it is enough to show that $\sum_{i=1}^m \beta_i^{1+q^0}$ is nonzero. Since $q$ is even, the characteristic of $\qm$ is 2.  Hence, $\sum_{i=1}^m \beta_i^{1+q^0}=\sum_{i=1}^m \beta_i^{2}=(\sum_{i=1}^m\beta_i)^2\neq 0$. Hence, if any $q$-ary image is self-orthogonal w.r.t the canonical inner product, then $\SC$ is self-orthogonal w.r.t the canonical inner product.
\end{proof}
\subsubsection{Self-dual basis}
Below is a well-known result. We give a novel proof using the ideas we have developed.
\begin{prop}\label{prop:selfdual}
Let $\SC$ be a scalable code over $\qm$, $\SB=\{\beta_1,\ldots,\beta_m\}$ be a basis of $\qm$ over $\q$ such that $\SB'=\SB$. $\im$ is self-orthogonal w.r.t the canonical inner product if and only if $\SC$ is self-orthogonal w.r.t the canonical inner product.
\end{prop}
\begin{proof}
Let $A$ be a matrix defined by
$$A= \left( \begin{array}{cccc} \beta_1 & \beta_1^q &\ldots &\beta_1^{q^{m-1}}\\
\beta_2 & \beta_2^q &\ldots &\beta_2^{q^{m-1}}\\
\vdots & \vdots &\vdots &\vdots \\
\beta_m & \beta_m^q &\ldots &\beta_m^{q^{m-1}} \end{array} \right ).$$
Since $\SB'=\SB$, we have Tr$(\beta_i\beta_j)=\delta_{ij}$ for $1\leq i,j \leq m$. Hence, $A\times A^T=I$, where $I$ is the $m\times m$ identity matrix and $A^T$ is the transpose of $A$. Hence, $A^T\times A=I$. The first row of $A^T\times A$ is $[\sum\beta_i^2,\ldots,\sum\beta_i^{1+q^{m-1}}]$. Hence, $\sum_{i=i}^m \beta_i^{1+q^k} = \delta_{0k}$ for $0\leq k\leq m-1$. From Theorem \ref{thm:IffImHerm}, it follows that $\im$ is self-orthogonal w.r.t the canonical inner product if and only if $\SC$ is self-orthogonal w.r.t the canonical inner product.
\end{proof}
\subsubsection{GF($q^2$) over GF($q$), $4|(q-1)$}
\begin{prop}\label{prop:1mod4quadcan}
Let $\SC$ be a scalable code over $\qq$, where $4|(q-1)$ and $\SB$ be a basis of $\qq$ over $\q$. If $\im$ is self-orthogonal w.r.t the canonical inner product, then so is $\SC$.
\end{prop}
\begin{proof}
From Theorem \ref{thm:IffImHerm}, it is enough to prove that for any basis $\{\alpha,\beta\}$, $\alpha^2+\beta^2\neq 0$. Let $\gamma$ be a primitive element of $\q$. Since $4|q-1$, $\gamma^{\frac{q-1}{4}}=i$ is a square-root of $-1$ and belongs to $\q$. Since $\alpha^2+\beta^2 = (\alpha+i\beta)(\alpha-i\beta)$ and $\{\alpha,\beta\}$ is a basis over $\q$ it follows that $\alpha^2+\beta^2\neq 0$ and we are done.
\end{proof}

It follows from Proposition \ref{prop:GcdEq} below that for the case of quadratic extensions, $\im$ being self-orthogonal forces $\SC$ to be self-orthogonal if and only if $q$ is even or $4|(q-1)$. Therefore, if $4|(q-3)$ one can have a non-self-orthogonal code $\SC$ such that $\im$ is self-orthogonal w.r.t the canonical inner product. Here is one possibility. 

\noindent {\it Example:} Consider self-orthogonality of images of codes from GF(9) over GF(3) w.r.t the canonical inner product. Let $\gamma$ be a primitive element of GF(9) such that $\gamma^2+\gamma+2=0$, $\gamma^8=1$ and $\gamma^4=-1$. The power sums of interest for a basis $\{\beta_1,\beta_2\}$ are $\beta^2_1+\beta^2_2$ and $\beta^4_1+\beta^4_2$. The basis $\SB=\{1,\gamma^2\}$ is such that $1+\gamma^4=0$ and $1+\gamma^8=-1$. Therefore, a scalable code $\SC$ self-orthogonal w.r.t the Hermitian-type product $\sum xy^3$ but non-self-orthogonal w.r.t the canonical inner product $\sum xy$ will result in an image (w.r.t the basis $\SB'$) that is self-orthogonal w.r.t the canonical inner product. Such a code can be easily constructed using the method given in Section \ref{sec:ExQuant}.

Finally, we remark that self-dual codes can be obtained as images of codes as well. Self-dual codes are linear codes which have rate half and are self-orthogonal w.r.t the canonical inner product. Since rate is preserved by imaging, image of a code is self-dual if and only if it is self-orthogonal w.r.t the canonical inner product and the original code has rate half. Like in the above example, it is possible to have a non-self-orthogonal, rate-1/2 code to result in a self-dual image, if the basis is chosen carefully.
\section{Quadratic Extensions}\label{sec:Quad}
We have seen before that if the trace of a code is self-orthogonal, all images are self-orthogonal. Converse is also true except in the case of binary images of 4-ary codes. This leads us to the search for situations where trace of a code is not self-orthogonal but an image with respect to some basis is self-orthogonal w.r.t a given Hermitian-type product. We begin by looking at quadratic extensions - i.e., GF$(q^2)$ over GF($q$). 

Let $q=p^r$, where $p$ is a prime number. Let $\SC$ be a scalable code of length $n$ over $\qq$ and $\SB$ be a basis of $\qq$ over $\q$ such that $\SB'=\{\alpha,\beta\}$. Let $f_{kl}$ be the Hermitian-type product as defined before. From Theorems \ref{thm:IffImHerm} and \ref{thm:IffTrHerm}, we know that self-orthogonality of $\im$ and $\tr$ w.r.t $\T{h}_l$ and $h_l$, respectively, is determined by self-orthogonality of $\SC$ w.r.t the forms $\sum_{i=1}^n x_iy_i^{p^l}$ and $\sum_{i=1}^n x_iy_i^{p^{l+r}}$ and the power sums $\alpha^{1+p^l}+\beta^{1+p^l}$ and $\alpha^{1+p^{l+r}}+\beta^{1+p^{l+r}}$. Here we would like to determine when these power sums can vanish and hence determine what self-orthogonality of $\im$  w.r.t $\T{h}_l$ implies about $\SC$.

Consider the power sum $\alpha^{1+p^l}+\beta^{1+p^l}$, where $0\leq l\leq 2r-1$ and $\{\alpha,\beta\}$ is a basis of $\qq$ over $\q$. This sum vanishes if and only if  there is a root of the equation $X^{1+p^l} + 1=0$ in $\qq$ which is not in $\q$, the root being $\frac{\alpha}{\beta}$. Hence, we would like to determine when every root of the equation $X^{1+p^l} + 1=0$ in $\qq$ is in $\q$. We distinguish two cases, viz. $p=2$ and $p$ odd.

\begin{prop}\label{prop:GcdEq}
Let $q=p^r$. Every root of the equation $X^{1+p^l} + 1=0$ in $\qq$ is in $\q$ - i.e., the power sum $\alpha^{1+p^l}+\beta^{1+p^l}$ does not vanish for any basis $\{\alpha,\beta\}$ of $\qq$ over $\q$, if and only if \\
(i) $p=2$ and gcd$(2^l+1,2^r+1)=1$ or \\
(ii) $p$ is odd and `` there is a power of two which divides $p^r-1$ but not $p^l+1$ and gcd$(p^l+1,p^r+1)=2$'' or ``every power of two dividing $p^{2r}-1$ divides $p^l+1$''.
\end{prop}
\begin{proof}
First consider the case $p=2$. There is a root of the equation $X^{1+2^l} +1 = 0$, say $\gamma$, in $\qq$ if and only if order of $\gamma$, which divides $2^{2r}-1$, also divides $1+2^l$.  Hence, there is a root of $X^{1+2^l}$ in $\qq$ if and only if gcd$(1+2^l,2^{2r}-1)>1$. $\gamma$ is in $\q$ if and only if order of $\gamma$ divides $2^r-1$. Hence, the following two statements are equivalent:\\
(i) Every root of the equation $X^{1+2^l} + 1=0$ in $\qq$ is in $\q$ \\
(ii) Every number dividing gcd$(1 +2^l,2^{2r}-1)$ divides $2^r-1$.\\
(ii) is clearly equal to the statement that gcd$(1+2^l,2^{2r}-1)|(2^r-1)$. Now, gcd$(2^r+1,2^r-1) = 1$ and $2^{2r}-1=(2^r-1)(2^r+1)$. Hence, gcd$(1+2^l,2^{2r}-1)|(2^r-1)$ if and only if gcd$(2^l+1,2^r+1)=1$. Hence, part (i) is true.

Suppose $p$ is odd. The equation $X^{1+p^l}+1=0$ has a root in $\qq$ if and only if there is an element whose order divides $2(1+p^l)$ and $p^{2r}-1$ but not $1+p^l$. This root is in $\q$ if and only if its order divides $p^r-1$. Hence, the following two statements are equivalent:\\
(i) Every root of the equation $X^{1+p^l} + 1=0$ in $\qq$ is in $\q$ \\
(ii) Every number dividing gcd$(2(1 +p^l),p^{2r}-1)$ but not $1+p^l$ divides $p^r-1$.\\
(ii) is clearly equivalent to the following statement:\\
(iii) gcd$(2(1+p^l),p^{2r}-1)|(p^r-1)$ or gcd$(2(1+p^l),p^{2r}-1)|(p^l+1)$

Let $p^l+1 = 2^a\prod_{i=1}^s p_i^{a_i}$, where $p_i$ are prime numbers and $a_i$ are non-negative numbers. We note that gcd$(p^r+1,p^r-1)=2$. Let $p^r+1 = 2^b\prod_{i=1}^t p_i^{b_i}$ and $p^r-1 = 2^c\prod_{i=t+1}^s p_i^{b_i}$, where $b_i$ are non-negative numbers. We have gcd$(2(1+p^l),p^{2r}-1) = 2^{\textrm{min}(1+a,b+c)}\prod_{i=1}^s p_i^{\textrm{min}(a_i,b_i)}$. 

Hence, gcd$(2(1+p^l),p^{2r}-1)|(p^r-1)$ if and only if min$(1+a,b+c)\leq c$ and min$(a_i,b_i)=0$ for $1\leq i\leq t$. min$(1+a,b+c)\leq c$ if and only if $a<c$. We know that $a\geq 1$. Since gcd$(p^r+1,p^r-1)=2$, $c\geq 2$ if and only if $b=1$.  Hence, min$(1+a,b+c)\leq c$ and min$(a_i,b_i)=0$ for $1\leq i\leq t$ if and only if $a<c$ and gcd$(p^l+1,p^r+1)=2$. Hence, gcd$(2(1+p^l),p^{2r}-1)|(p^r-1)$ if and only if there is a power of two which divides $p^r-1$ but not $p^l+1$ and gcd$(p^l+1,p^r+1)=2$. 

gcd$(2(1+p^l),p^{2r}-1)|(p^l+1)$ if and only min$(a+1,b+c)\leq a$ - i.e., if and only if $b+c\leq a$ - i.e., every power of two dividing $p^{2r}-1$ divides $p^l+1$. Hence, part (ii) is true.
\end{proof}

\begin{prop}\label{prop:2DivrNotl}
Let $q=p^r$ and $l\neq 0$. Every root of the equation $X^{1+p^l} + 1=0$ in $\qq$ is in $\q$- i.e., the power sum $\alpha^{1+p^l}+\beta^{1+p^l}$ does not vanish for any basis $\{\alpha,\beta\}$ of $\qq$ over $\q$, if there is a power of two which divides $r$ but not $l$.
\end{prop}
\begin{proof}
Suppose $p=2$. From the proof of Proposition \ref{prop:GcdEq}, every root of the equation $X^{1+p^l} + 1=0$ in $\qq$ is in $\q$ if gcd$(1+2^l,2^{2r}-1)|(2^r-1)$. Clearly, gcd$(1+2^l,2^{2r}-1)|\textrm{gcd}(2^{2l}-1,2^{2r}-1) = 2^{2\textrm{gcd}(l,r)}-1$. Additionally, $2^{2\textrm{gcd}(l,r)}-1|2^r-1$ if and only if there is a power of two which divides $r$ but not $l$. Hence, the result is true.

Suppose $p$ is odd. From the proof of Proposition \ref{prop:GcdEq}, every root of the equation $X^{1+p^l} + 1=0$ in $\qq$ is in $\q$ if gcd$(2(1+p^l),p^{2r}-1)|(p^r-1)$. Since $(p^l-1)/2$ is an integer, gcd$(2(1+p^l),p^{2r}-1)|$gcd$(p^{2l}-1,p^{2r}-1) = p^{2\textrm{gcd}(l,r)}-1$. Additionally, $p^{2\textrm{gcd}(l,r)}-1|p^r-1$ if and only if there is a power of two which divides $r$ but not $l$. Hence, the result is true.
\end{proof}

Let $\T{h}_l$ and $h_l$ be the Hermitian-type products as defined in the previous section. Proposition \ref{prop:2DivrNotl} immediately leads to the following two results:
\begin{cor}\label{cor:ImTr2DivrNotl}
Let $q=p^r$ and $l\neq 0$. Let $\SC$ be a scalable code over $\qq$ and $\SB$ be a basis of $\qq$ over $\q$. If there is a power of two which divides $r$ but not $l$, then $\im$ is self-orthogonal w.r.t $\T{h}_l$ if and only if $\tr$ is self-orthogonal w.r.t $h_l$.
\end{cor}
\begin{proof}
By Theorems \ref{thm:IffImHerm} and \ref{thm:IffTrHerm} and the discussion in the starting of this section, self-orthogonality of $\im$ w.r.t $\T{h}_l$ and $\tr$ w.r.t $h_l$ are equivalent if and only if every root of the equations $X^{1+p^l} + 1=0$ and $X^{1+p^{l+r}} + 1=0$ in $\qq$ is in $\q$. By Proposition \ref{prop:2DivrNotl}, this is possible if there is a power of two which divides $r$ but not $l$ and $r+l$ which is possible if and only if there is a power of two which divides $r$ but not $l$. Hence, the result follows.
\end{proof}

\begin{cor}\label{cor:ImTrrPow2}
Let $q=p^r$. Let $\SC$ be a scalable code over $\qq$ and $\SB$ be a basis of $\qq$ over $\q$. If $r$ is a power of two, then $\im$ is self-orthogonal w.r.t $\T{h}_l$ if and only if $\tr$ is self-orthogonal w.r.t $h_l$ for $1\leq l\leq r-1$ and $r+1\leq l\leq 2r-1$.
\end{cor}
\begin{proof}
If $r$ is a power of two and $1\leq l\leq r-1$ and $r+1\leq l\leq 2r-1$, then there is a power of two which divides $r$ but not $l$, the power being $r$ itself. Hence, the result follows from Corollary \ref{cor:ImTr2DivrNotl}.
\end{proof}

From Proposition \ref{prop:GcdEq}, we see that studying the behavior of gcd$(p^r+1,p^l+1)$ is beneficial. Suppose that $r\geq l$. Then $r$ can be written as $r=al+b$, where $0\leq b<l$. Hence, gcd$(p^r+1,p^l+1)$ = gcd$(p^r-p^l,p^l+1)$ = gcd$(p^{r-l}-1,p^l+1)$ = gcd$(p^{r-l}+p^l,p^l+1)$ = gcd$(p^{r-2l}+1,p^l+1)=\ldots=$ gcd$(p^b+(-1)^a,p^l+1)$. Similarly we see that the following results are true:
\begin{eqnarray*}
\textrm{gcd}(p^{al+b}+1,p^l+1) &=& \textrm{gcd}(p^b+(-1)^a,p^l+1)\\
\textrm{gcd}(p^{al+b}-1,p^l+1) &=& \textrm{gcd}(p^b-(-1)^a,p^l+1)\\
\textrm{gcd}(p^{al+b}+1,p^l-1) &=& \textrm{gcd}(p^b+1,p^l-1)\\
\textrm{gcd}(p^{al+b}-1,p^l-1) &=& \textrm{gcd}(p^b-1,p^l-1).
\end{eqnarray*}
From this it follows that gcd$(p^r\pm1,p^l\pm1)$ takes one of these four values: $1, 2, p^{\textrm{gcd}(r,l)}+1, p^{\textrm{gcd}(r,l)}-1$. Hence, just by computing gcd$(l,r)$ and checking for divisibility we can compute the values of gcd$(p^r\pm1,p^l\pm1)$.

Finally, we note that the results relating to power sums which have been derived in this section can be used to determine what self-orthogonality of $\im$ w.r.t $\T{f}$ implies about $\SC$.
\section{Quantum Code Construction}\label{sec:ExQuant}
In this section, we specialize our results to cyclic codes and construct new quantum BCH codes from 4-ary images of $4^m$-ary codes. Suppose $\SC$ is a cyclic code of length $n$ over $\qm$ with generator polynomial $g(x) = \prod_{i\in Z} (x-\alpha^i)$, where $\alpha$ is a primitive $n$th root. Then the set $Z$ is called the \emph{zeros of the code} and its complement $S$ is called the \emph{nonzeros of the code}. In our examples, we consider cyclic codes with blocklength $n|(q^m-1)$; therefore, the zero set can be any subset of $\{0,1,\ldots,n-1\}$. In some cases, the codes happen to be Reed-Solomon (RS) codes. The following two propositions are standard results about cyclic codes \cite{sloane} that are used in our construction. We provide short proofs for completeness.

\begin{prop}\label{prop:TrZeroSet}
Let $\SC$ be a cyclic code of length $n$ over $\qm$ with zero set $Z$ and non-zero set $S$. For $0\leq s\leq n-1$, let $C_s$ denote the cyclotomic coset modulo $n$ under multiplication by $q$ containing $s$. Then $\tr$ has non-zero set $S^c=\cup_{s\in S}C_s$ and zero set $Z^c=\cup_{\{s|C_s\subseteq Z\}}C_s$.
\end{prop}
\begin{proof}
If $\SC$ has zero set $Z$ and non-zero set $S$, then the subfield subcode $\SC|\q$ has zero set $\cup_{s\in Z}C_s$. By Delsarte's theorem \cite{sloane}, $\tr$ $= (\SC^\perp|\q)^\perp$. Hence, $\tr$ has non-zero set $-\cup_{s\in -S}C_s = \cup_{s \in S} C_s =S^c$ and so $\tr$ has zero set  $\cup_{\{s|C_s\subseteq Z\}}C_s = Z^c$.
\end{proof}

\begin{prop}\label{prop:CyclicHerm}
Let $\SC$ be a cyclic code of length $n$ over $\qm$ with zero set $Z$ and non-zero set $S$. Then the following are equivalent:\\
(1)$\SC$ is self-orthogonal w.r.t the form $\sum x_i y_i^{p^l}$\\
(2)$(-p^l S)(\textrm{mod } n) \subseteq Z$\\
(3)$(-p^{-l} S)(\textrm{mod } n) \subseteq Z$
\end{prop}
\begin{proof}
Let $\SC' = \{ (x_1^{p^l},\ldots,x_n^{p^l}):(x_1,\ldots,x_n)\in\SC\}$. $\SC'$ has zero set $(p^l Z)(\textrm{mod } n)$ and non-zero set $(p^l S)(\textrm{mod } n)$. $\SC$ is self-orthogonal w.r.t the form $\sum x_i y_i^{p^l}$ if and only if $\SC'\subseteq \SC^\perp$, which is equivalent to the condition $(p^l Z)(\textrm{mod } n)\supseteq -S$. Taking complements, we have $(1)\Leftrightarrow (2)$. Dividing both sides by $p^l(\textrm{mod } n)$, we have $(1)\Leftrightarrow (3)$
\end{proof}

We now consider some examples of codes which can be used to generate quantum codes. Consider cyclic codes of length $n$ over GF$(4^m)$ with zero set $Z$ and non-zero set $S$. Let $\SB$ be a basis of $\qm$ over $\q$ and $\SB'=\{\beta_1,\ldots,\beta_m\}$ be the dual basis of $\SB$. From Propositions \ref{prop:TrZeroSet} and \ref{prop:CyclicHerm} and Theorems \ref{thm:IffImHerm} and \ref{thm:IffTrHerm}, $\tr$ is self-orthogonal w.r.t the Hermitian inner product if and only if
$$-2S^c (\textrm{ mod }n)\subseteq Z^c,$$ 
and $\im$ is self-orthogonal w.r.t the Hermitian inner product if and only if
\begin{eqnarray*}
&-2^{2k+1}S \textrm{ mod }n \subseteq Z\\
&\text{ for } k\in\{0,1,\ldots,m-1\} \text{ such that } \sum_{i=1}^{m}\beta_{i}^{1+2^{2k+1}}\ne0.
\end{eqnarray*}
From the BCH bound, the minimum distance of $\SC$ and $\SC^\perp$ is at least 1 greater than the number of consecutive integers in $Z$ and $S$, respectively. 

\noindent {\it Example:} Consider GF(16) over GF(4). Here $q=4=2^2$ and $l=1$. From Corollary \ref{cor:ImTrrPow2}, we know that there can be no scalable code whose image is self-orthogonal w.r.t the Hermitian inner product but not trace. Hence, in this case, there can be no improvement over the quantum codes given in \cite{andrew}.

\noindent {\it Example:} Consider GF(64) over GF(4). Let $\alpha$ be a primitive root of the polynomial $X^6+X+1$ in GF(64). The power sums of interest in a dual basis $\{\beta_1,\beta_2,\beta_3\}$ are $\beta^3_1+\beta^3_2+\beta^3_3$, $\beta_{1}^{9}+\beta_{2}^{9}+\beta_{3}^{9}$, and $\beta_{1}^{33}+\beta_{2}^{33}+\beta_{3}^{33}$. 
\begin{enumerate}
\item Let $n=63$. $\SB_2=\{1,\alpha,\alpha^5\}$ is a basis such that the sum of 9th powers is zero. Hence, $S\subseteq \{1,2,\ldots,62\}$ such that $(-2S\cup -32S)\subseteq Z$ and $-2S^c\nsubseteq Z^c$ leads to a cyclic code whose image w.r.t $\SB'_2$ is self-orthogonal but not trace. An example is $S=\{1,2,\ldots,20\}$. This code leads to an [[189,69,21]] quantum code and has largest minimum distance among quantum codes of length 189 obtained by images of cyclic codes of length 63 over GF(64). The table of codes from \cite{andrew} shows that trace is self-orthogonal for codes with nonzero sets $\{1\}$ to $\{1,2,3,4,5,6\}$. Hence, the maximum minimum distance possible was limited to 7 for trace-self-orthogonal codes. Using self-orthogonality of images has resulted in the possibility of codes with minimum distance up to 21.
\item Let $n=7$. $\SB_1=\{1,\alpha^3,\alpha^{15}\}$ is a basis such that the sum of 3rd and 33rd powers is zero. $S=\{1,2,3\}$ is such that $-8S= \{4,5,6\}, S^c=\{1,2,3,4,5,6\},$ and $-2S^c = S^c$. Hence, its image w.r.t $\SB'_1$ is self-orthogonal but not trace. This code leads to an [[21,3,4]] quantum code and has largest minimum distance among quantum codes of length 21 obtained by images of cyclic codes of length 7 over GF(64).
\end{enumerate}

Table \ref{Tab:Quant} is a partial list of quantum codes obtained by taking 4-ary images of cyclic codes over GF(16) and GF(64).
\begin{table}[!tbp]
\center
\caption{Parameters $[[n,k,d]]$ of quantum codes for $m= 2,3$ and $n_0 =15,7,63$. $S$ is the nonzero set of the cyclic code over  $\textrm{GF}(4^m)$. $n=mn_0, k = n - 2m|S|, d = |S|+1$. Notation for basis is from examples.}
\label{Tab:Quant}
\begin{tabular}{|cc|ccc|c|c|}
\hline
m&$n_0$&n&k&d&S&Basis\\
\hline
2 &15&30&26&2&\{1\}&All\\
&&30&22&3&\{1,2\}&All\\
&&30&18&4&\{1,2,3\}&All\\
&&30&14&5&\{1,2,3,4\}&All\\
\hline
3&7&21&15&2&\{1\}&All\\
&&21&9&3&\{1,2\}&All\\
&&21&3&4&\{1,2,3\}&$\SB'_1$\\
\hline
3&63&189&183&2&\{1\}&All\\
&&189&177&3&\{1,2\}&All\\
&&189&171&4&\{1,2,3\}&All\\
&&189&165&5&\{1,2,3,4\}&All\\
&&189&159&6&\{1,2,3,4,5\}&All\\
&&189&153&7&\{1,2,3,4,5,6\}&All\\
&&189&147&8&\{1,2,3,4,5,6,7\}&$\SB'_2$\\
&&189&141&9&\{1,2,3,4,5,6,7,8\}&$\SB'_2$\\
&&\vdots &\vdots &\vdots &\vdots&\vdots\\
&&189&75&20&\{1,2,3,\ldots,18,19\}&$\SB'_2$\\
&&189&69&21&\{1,2,3,\ldots,18,19,20\}&$\SB'_2$\\
\hline
\end{tabular}
\end{table}
\section{Conclusion}\label{sec:conc}
We have derived necessary and sufficient conditions for self-orthogonality of images of codes with respect to a general biadditive form. The conditions separate into a power sum criterion on the dual basis elements and self-orthogonality of the original code with respect to conjugate biadditive forms. The condition can be easily applied to practical codes such as cyclic codes to construct self-orthogonal codes. We have derived several interesting corollaries to the main result and showed a possible application in the construction of quantum codes.

Several avenues for future work are possible. The case of quadratic extensions and Hermitian-type products has been studied in detail. In particular, we have been able to find many cases for which self-orthogonality of an image is possible only through the self-orthogonality of the trace. An interesting problem is to extend this study to images of codes from GF($q^m$) over GF($q$) for $m\ge3$. Can there be situations where self-orthogonality of an image implies self-orthogonality of the trace for $m\ge3$? The answer could probably be obtained through the study of power sums of basis elements. 

\end{document}